# Software Defined Networking Supported by IEEE 802.1Q

János Farkas, Stephen Haddock, Panagiotis Saltsidis
janos.farkas@ericsson.com, shaddock@stanfordalumni.org, panagiotis.saltsidis@ericsson.com

*Abstract*— **This paper gives an overview on how Software Defined Networking (SDN) principles can be applied to existing Ethernet merchant silicon considering the requirements modern networks face. We show that existing Layer 2 features specified by IEEE 802.1Q support SDN. The bridge architecture [1] supports control plane/data plane split by design and also allows for external control e.g. by an SDN Controller. The data plane provided by existing chips is feature rich for network virtualization and supports even more features like OAM. We outline the principles of SDN over bridges and show a number of possibilities for further research and development.**

I. INTRODUCTION

Software Defined Networking (SDN) is an emerging new networking paradigm, which aims to introduce a new approach to the control and design of networks of various kinds. SDN relies on directly programming the packet handling mechanisms of the network nodes by a network controller. That is, the SDN concept allows defining the networking behavior via software tools that are easy to modify as opposed to behavior hard-coded in the equipment by design. It is understood that the behavior of the networking equipment is defined by software today; however, it is often difficult to change the behavior and requires expert knowledge of the equipment. As opposed, SDN provides flexibility along the following three characteristic features:

- Programmability of the network
- Separation of the control plane from the data plane
- A controller that has a view of the entire network and can control the network devices

There are a number of ways to implement SDN programming of the network nodes. There are some approaches designed with SDN in mind, such as ForCES [9], OpenFlow [14], OpenDaylight [15] or OpenStack [16]. There are other approaches where the original goal was not SDN but they can be applied for SDN as well, e.g. SNMP or NETCONF. The exact implementation choice is out of scope for this paper. Instead, we focus on the principles and their application on a high level. We note that the term "distributed control" is used in this paper to denote existing distributed bridge control protocols.

Similar to networks under distributed control, SDN networks have to meet several requirements, which depend on the exact application. SDN addresses various network scenarios from enterprise through campus to carrier grade networks, including data center and backhaul networks. Therefore, prospective SDN approaches have to meet a large part of the below requirements:

- Providing L2 and L3 connectivity services
- Network virtualization
- Supporting several customers or tenants
- Scalability
- Decoupling logical and physical configuration
- Address separation
- Traffic isolation
- Supporting station mobility, e.g. virtual machine (VM) mobility
- Quality of Service (QoS) assurance
- Auto-provisioning and service discovery
- Operations, Administration and Maintenance (OAM)

which are discussed more in detail in the following. Most of these requirements appear whether network virtualization is provided based on L2 [1] or based on L3 [12].

Hosts, servers, network devices and their virtual equivalents are often (if not exclusively) reached via either Internet Protocol (IP) or Ethernet for communication. Therefore, Layer 2 and Layer 3 connectivity is an essential requirement.

Network virtualization is crucial for a network or a cloud provider in several ways and represents a high level requirement that is related to several of the remaining requirements. For example, scalability – among other things- requires the support of a number of customers or tenants. Similarly, scalability requires that physical resources may be re-used across multiple customers. This is an essential characteristic of network virtualization and depends on the ability to decouple logical networks from the physical network as well as the separation of address spaces, which are necessary to eliminate address assignment dependencies among the customers and for the support of mobility, e.g. VM mobility. In order to prevent mis-delivery of one customer's data to other customers, traffic for each logical network (or virtual network instance) must be isolated from traffic of all other virtual network instances. Scalability is also a general requirement in order (for instance) to avoid explosive growth in forwarding table size. Creation of a new virtual network instance or modification of an existing one (such as expansion) must be something that the provider can do with relative ease. Difficulty in changing a service impacts on the ability of the provider to increase revenue and/or control the expenses associated with service provisioning especially in a cloud environment. Auto-provisioning and verification of a virtual network must be enabled by network features. For services with traffic classes requiring differentiated Quality of Service (QoS), this must be enabled by the network. In order to support monitoring and diagnosis of services, a solution must support Operations Administration and Maintenance (OAM) capabilities on a per-virtual network (per service) basis.

This is a lot of requirements. The good news for early SDN deployments is that there is no need to do it from scratch. Actually, some networking technologies already provide a very good basis for SDN, which could be leveraged by SDN research and development. This paper explores Ethernet networking, i.e. bridging features useful for SDN. The reason to focus on Ethernet is its key role in networking today. The paper shows that the existing Ethernet data plane features form a good basis for SDN systems. Furthermore, the bridging standard [1] supports the separation of the control plane from the data plane by design and also allows control by an SDN Controller. Additional features, such as OAM or VM migration, are not only described by the standard, but are already supported by merchant silicon today. The paper also demonstrates how Ethernet meshes with SDN and outlines the design principles of a potential Ethernet-based SDN system.

The rest of the paper is structured as follows. Section II explains the architecture. Section III provides details on network control. Network virtualization is then explained in Section IV. The puzzles are then put into a big picture in Section V by presenting a hybrid network approach and by providing a network example in Section VI. The paper is finally summarized in Section VII.

## II. SDN AND BRIDGE ARCHITECTURE

Splitting the architecture into separate control and data planes is beneficial for independent scaling and innovation. It allows modularity and helps fractioning the functionality into components that can be well defined.

### A. SDN Architecture

The control and data planes are separated in the SDN architecture as shown in Fig. 1. The data plane is configured with the blueprint of the actions invoked on an incoming packet out of the possible Action Set supported by the device.

The control plane has two components: local control and remote control. Typically, there is a central remote controller that performs all the computation required for determining the forwarding paths for the data packets. Furthermore, the SDN Controller determines how the network nodes should be set up in order to achieve the desired forwarding behavior, i.e. exactly which actions are invoked on the packets out of the commands available to the network nodes by means of an SDN Protocol. The programming of the network is then implemented by the local control in the nodes, which interprets the instructions from the remote control and configures the local data plane accordingly.

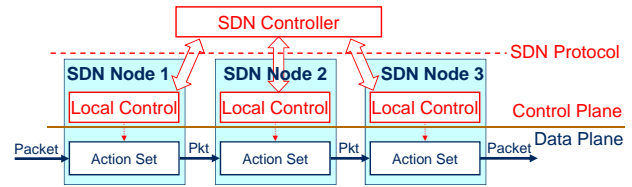

Fig. 1. SDN Architecture

Overall, the data plane consists of generic boxes programmed by the control plane such that the network behavior is determined by the remote SDN Controller controlling several devices.

### B. Bridge Architecture

IEEE 802.1 standards rely on a model that is based on a clear separation of the data and the control planes. This fact is often overlooked because the control plane was originally distributed and the separation was inside the bridges. The MAC bridge architecture is specified by IEEE 802.1Q [1][1]; and is illustrated in Fig. 2. Note that MAC bridges are often referred to as Ethernet switches because IEEE 802.3 Ethernet is the most common IEEE 802.n media access method for IEEE 802.1 bridges. The distributed control protocols, e.g. Shortest Path Bridging (SPB) [3], are implemented by the so-called Higher Layer Entities, which then control the data plane as shown in the figure. In addition, the standard also allows control by an External Agent [4], even co-exist with distributed control in the same network. Distributed control is turned off for the packets controlled by External Agents.

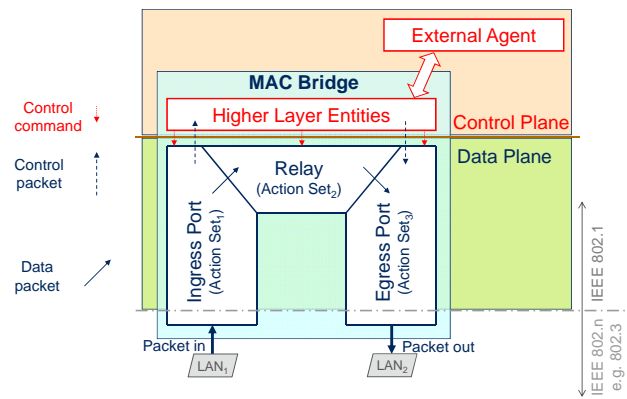

Fig. 2. Bridge architecture

---

[1] When referring to "IEEE 802.1Q" one actually refers to the latest approved revision of the standard, which is 802.1Q-2011 today. Note that a revision [2] is ongoing, which e.g. merges [3] and [6] into the main standard.

External Agents providing topological separation to the already split control and data planes were introduced by 802.1Qay [4], which has been already merged to the bridge standard [1]. The External Agent can for example be an SDN Controller, a Path Computation Element (PCE) [10] or even a protocol like the Generalized Multiprotocol Label Switching (GMPLS) [11]. If the control is provided by one or more External Agents, then the task of the control (Higher Layer Entity) local to the bridge is to implement the instructions of the External Agent. The standard [1] specifies the Information Model and the Data Model that External Agents can rely on.

The data plane of a bridge shown in Fig. 2 depicts two ports and a relay in-between them. For the ease of explanation, the ports have a direction in the figure, i.e. Ingress and Egress represent their role for a single packet. Data packets are received by the Ingress Port, which may perform one or more actions on the packet out of its Action $Set_1$, depending on how it is programmed. Data packets are then sent to the central processing, i.e. to the Relay, which can also perform actions out of its Action $Set_2$. Finally, the Egress Port carries out actions from Action $Set_3$. Control packets are sent to Higher Layer Entities by the Ingress Port and the Egress Port may receive them from Higher Layer Entities as shown in the figure.

It is either the External Agent or the distributed control that determines what exactly happens to a data packet. The actions are grouped into three sets: ingress, relay and egress action sets. Each action set of a standard bridge provides a wide range of programmable features, which are discussed in the following.

Action $Set_1$ of an Ingress Port of a bridge involves the following actions that can be performed on a data packet:

- Drop (filter)
- Tagging, untagging
- Virtual LAN (VLAN) IDentifier (VID) translation
- Encapsulation, decapsulation
- Metering

The packet is dropped if Ingress Filtering is turned on and the Ingress Port is not a member of the VLAN the packet belongs to (based on the outermost VLAN tag carried in the packet). In addition, the packet may also be dropped for loop mitigation. Furthermore, the Ingress Port may add a new tag or a new Ethernet header to the packet as outermost header fields; or may remove the outermost tag or header. In addition, VID translation can also be performed based on the VID translation table, i.e. the outermost VID can be replaced by another VID. Metering may result in marking or dropping packets exceeding bandwidth limits.

The Relay is responsible for forwarding the packet to output ports based on the VLAN ID and the destination address carried in the packet. The operation of the Relay is based on forwarding tables, which may contain entries of various types. The Relay may also drop the packet. That is Action $Set_2$ is either forward or drop based on table entries.

Action $Set_3$ of Egress Ports involves the following actions:

- Drop (filter)
- Tagging, untagging
- VID translation
- Encapsulation, decapsulation
- Queueing
- Transmission selection

That is, the Egress Port drops the packet if Egress Filtering is turned on and the port is not member of the VLAN that the packet belongs to. The Egress Port may remove or add an outer tag or header. VID translation may be also performed based on the VID translation table. Queuing and transmission selection governs how a packet is sent out.

In summary, the bridge architecture provides a wide range of knobs for network programmability. The bridge architecture splits the control plane from the data plane; furthermore, it allows network control or programming by an external entity thus allowing geographical separation. That is, the bridge architecture provided by the standard [1] is in-line with the three main characteristics of SDN.

III. NETWORK CONTROL

Ethernet networks have multiple topology layers, which are shown in Fig. 3. All these layers can be programmed by SDN or controlled by an appropriate distributed control protocol as illustrated in the figure. The control of the network is based on the control of these layers; therefore, it is discussed in this section. The topology layers lay down the basics for network virtualization, which is discussed in the next section.

The physical topology is the bottom layer, which is managed by means of enabling or disabling the ports of the nodes by SDN control or by the network management.

On top of the physical topology, there is the loop-free active topology, which is a subset of the physical topology and contains the active links. The active topology is comprised of trees; shortest path trees, spanning trees or explicit trees. Aside from SDN control, the active topology can be controlled by IS-IS or by a spanning tree protocol. An SDN Controller may even leverage the Intermediate System to Intermediate System (IS-IS) [5] routing protocol in order to easily maintain explicit trees [8].

The VLAN topology is practically determined by the VLAN membership of the ports of the nodes, which is typically a subset of the active topology. The VLAN membership of the ports can be controlled by SDN aside distributed control.

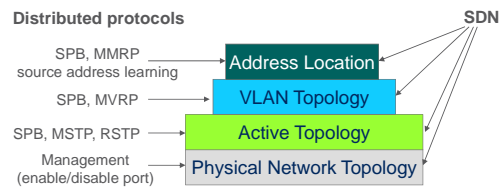

Fig. 3. Topology layers in an Ethernet LAN

The forwarding table entries, i.e. the location of the addresses destined by the data traffic also define a topology within a VLAN. The sum of the forwarding paths to a unicast destination form a tree rooted at the destination. In case of distributed control, forwarding table population can be performed e.g. by SPB. Furthermore, MAC auto learning, i.e. learning of the source addresses of data packets may be also performed. Naturally, table entries can also be manipulated by External Agents – a hook for SDN. MAC learning from data packets can be turned off for external control and for SPB.

As we can see the different topology layers provide very flexible control of the forwarding paths with several knobs to manipulate them. Let us investigate the control options a bit more in detail.

*A. SDN Control*

An SDN Controller (i.e. an External Agent) can program each topology layer shown in Fig. 3. That is, it is fully up to the controller to determine which manner to program the different topology layers in order to achieve the desired forwarding behavior if the network is under SDN control. For example, VLAN membership of ports can be set and forwarding table entries can be inserted or removed by the SDN Controller. This, in effect will result in controlling filtering and encapsulation behavior of the switches and can potentially lead to new behavior.

SDN based on Ethernet can be implemented for example as follows. As Section II explains, a standard data plane model is defined for Ethernet, which involves the packet fields, the sets of actions and their compositions. The standard [1] defines the Information Model and the Data Model that the SDN Protocol shown in Fig. 1 can use for controlling the bridges by an SDN Controller. During the specification of 802.1Qay [4], SNMP was considered as the SDN Protocol. For controlling explicit trees and paths, IS-IS [8] can be used as the protocol for instructing the bridges by an SDN Controller. Other SDN Protocols can be used for the programming of the data plane if for example the Local Control shown in Fig. 1 performs translation between the SDN Protocol and the models specified by the standard [1]. Even though each chip implements a proprietary API to manipulate the data plane, they provide access to program the data plane by the SDN Protocol as they are compliant to the standard.

*B. Distributed Control*

SPB is considered as the main form of distributed control in this paper because it is able to control all the topology layers shown in Fig. 3 except for the physical topology. SPB is based on the Intermediate System to Intermediate System (IS-IS) [5] link state routing protocol. As such, SPB has an in-built auto-discovery for topology and for the services or addresses assigned to network nodes. SPB then automatically sets the forwarding paths necessary to provide the connectivity based on its auto-discovery.

The control protocol for a particular VLAN can be selected by allocating the VLAN to the Multiple Spanning Tree Instance (MSTI) dedicated to the control protocol aimed to be used. There is an MSTI dedicated to External Agents, which is referred to as Ext-MSTI (hex FFE) [4] in the following. The rest of the MSTIs are under distributed control as specified today, e.g. three MSTIs are associated with IS-IS control. VLANs that are not touched by distributed control but controlled by an SDN Controller have to be allocated to the Ext-MSTI.

Taking a look on the network requirements listed in the introduction, we can see that a couple of them are already met by the features discussed up to this point. The auto-discovery of link state SPB provides service discovery, which is explained more in detail in Section VI. Furthermore, the auto-discovery supports station migration and the mapping of addresses to services and to VLAN tunnels.

IV. NETWORK VIRTUALIZATION

As the previous section explained, basic Ethernet already provides network virtualization by means of Virtual LANs. The specialty of this virtualization is that the ID of the virtual network is carried in the header of data packets thus making possible to decide which virtual network the packet belongs to. This makes the provisioning of virtual networks easy. Nevertheless, its scalability was limited by the 12-bit VID space. Therefore, further virtualization techniques have been added to Ethernet, thus scalability limitations have been resolved. Fig. 4 depicts all the possible Ethernet header formats available today for network virtualization.

The widely-known VLAN tagging standardized in 1998 is referred to as Customer VLAN (C-VLAN) tagging (second column). The next step was the specification of the Service VLAN (S-VLAN) tag introduced by Provider Bridges (PB) [1] which is sometimes referred to as Q-in-Q, due the use of two VLAN tags. Thus, instead of the former 12 bits, 24 bits were provided for network virtualization. After that, full Ethernet header encapsulation was introduced by Provider Backbone Bridges (PBB) [1], which is sometimes referred to as MAC-in-MAC, due to the encapsulation in another full MAC header.

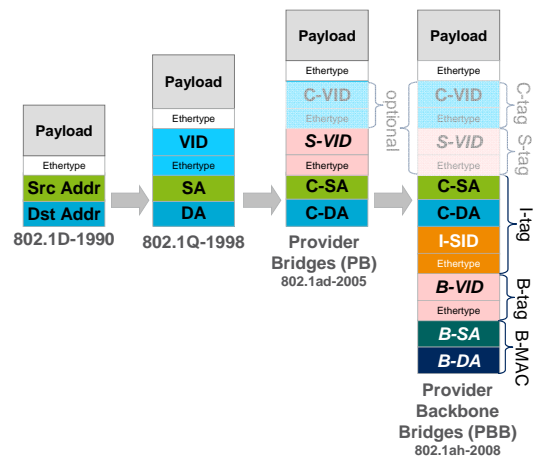

Fig. 4. Encapsulation formats provided by Ethernet

Besides keeping the formerly specified VLAN tags, a new 24-bit ID referred to as I-SID was introduced for service identification, which in fact provides a 24-bit Layer 2 virtual network ID. This means that 16 million virtual networks can be supported by one Backbone VLAN (B-VLAN). Altogether a 60-bit space is provided for network virtualization by Ethernet today, which removes all scalability concerns. Note, that in case of PBB, the payload with its Ethertype may immediately follow the I-tag or can optionally be preceded by VLAN tags as illustrated in the figure. Let us now take a look on how the Ethernet packet formats provide network virtualization.

In effect, network virtualization means providing overlay networks. A basic overlay is provided by the C-VLAN on top of the physical topology as shown in Fig. 3. Interpreting the overlay as a service provided by the network, the VID is the service ID, which also has a key role in the forwarding of the data packet, i.e. in the transport. The full PBB packet format allows 4 layers of overlays as shown in Fig. 5. Each of these overlays may be used for example if customer networks are connected to PB networks connected to Edge Bridges (EB) of a provider's PBB network or in a PBB Data Center Network (DCN) as illustrated in the figure.

A key aspect of PBB is that it separates the service layer from the transport layer. That is, the first overlay provided by the physical network is the B-VLAN for the transport and the service layer is on top of the B-VLANs. A service provided by the backbone is identified by an I-SID, which can be point-to-point, multipoint-to-multipoint or rooted multipoint. An I-SID can offer different services, i.e. further overlays. In the example shown in Fig. 5, the I-SID provides an S-VLAN overlay, which then provides a C-VLAN overlay. All of the overlays, i.e. virtual networks can be controlled by SDN.

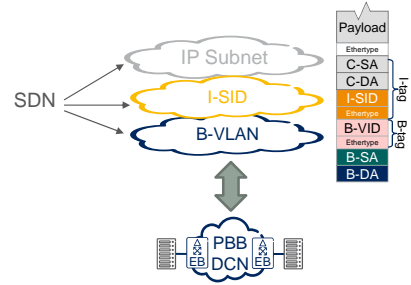

Fig. 6. Layer 3 overlay

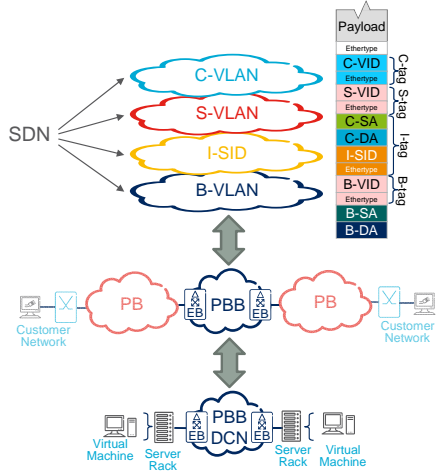

Fig. 5. Layer 2 overlays

IP is a native overlay for Ethernet according to the ISO layering. It is quite common to associate an IP subnet to a VLAN. In case of PBB, the overlay service provided by the I-SID can be a Layer 3 Virtual Private Network (L3VPN) as illustrated in the DCN example of Fig. 6. In this case, the optional fields of the PBB header are not present. More details on L3 overlays are available e.g. in [13].

Flow-based programmability is provided for network wide SDN by means of mapping data flows to a flow ID at edge bridges and programming the forwarding for the flow ID throughout the network. That is, the mapping of data flows to I-SIDs or B-VIDs can be much more flexible than the direct mapping of an IP subnet or a VLAN to an I-SID, and the direct mapping of an I-SID to a B-VID as discussed before. In fact, arbitrary mapping can be applied on the data flows in the edge nodes of the networks. A flow, for example, can be mapped to an I-SID, a B-VID, or a Flow Hash [7] based on an n-tuple classification or any other field in the packet, e.g. TCP port. Having the classification and the per-flow mapping implemented by the edge bridges, core bridges can perform the forwarding based on the standard Ethernet header without performing deeper packet inspection.

As discussed in this section, an SDN controlled network providing L2 and L3 connectivity leveraging the existing Ethernet features is able to provide network virtualization fulfilling all network virtualization related requirements. This means that a large number of tenants or customers can be supported due to the scalability provided. Furthermore, the customer or tenant may have its own customers because of the several layers of virtual networks provided.

## V. HYBRID NETWORKS

There are a couple of features already available in Ethernet, today but difficult to do in a software defined manner (or at least not part of centralized SDN solutions). Among these, OAM and fast protection switching are the most important ones for carrier-grade networks.

Hybrid networks are comprised of hybrid nodes that support both SDN and distributed control. The hybrid use of SDN and distributed control enables using the existing features and make them available for SDN right now almost for free. Moreover, SDN bootstrapping can rely on distributed control (e.g. IS-IS) in a hybrid network, which ensures a default in-band control channel for the SDN Protocol. Some further advantages of hybrid networks are discussed in the following.

In order to be able to use existing chips and avoid the need for complete replacement of each network node and host, packet formats should not be changed by SDN, at least not initially. This even allows the data plane interworking of devices controlled along different principles, i.e. one can be under distributed control the other one can be under centralized SDN control; which also provides a smooth migration path. As a result, hybrid operation becomes just the matter of proper control, i.e. carefully crafted co-existence and operation of SDN and distributed control in the same network. In case of Layer 2, special attention is required to preserve the strict loop-free operation of existing distributed control, since loops can cause network meltdowns.

A way for achieving the desired proper coexistence of SDN and distributed control is already supported by the bridging standard [1], which is based on VLAN separation. The clear split of the VLAN space and assigning the VLAN sets to the desired control planes ensures proper operation for both the SDN and the distributed control thus avoiding any state conflict or ambiguity in the operation. The standard ensures that a forwarding table can be only controlled by a single control plane, thus forwarding table separation is also provided besides VLAN separation. The control protocol operation mode for a VLAN can be selected by allocating the VLAN to the MSTI associated with the desired operation mode as discussed before in Section III.B. SDN VLANs and forwarding tables are allocated to the Ext-MSTI, hence the SDN Controller sets up the forwarding paths for these VLANs. Distributed control is completely turned off for the Ext-MSTI. The clean separation ensures that conflict is not possible between SDN and the distributed control.

Besides carrier-grade networks, OAM tools are essential for the maintenance of most networks. It is critical to ensure fate sharing between data and OAM packets. In order to achieve fate sharing, the operation of Connectivity Fault Management (CFM), which is the Ethernet OAM, relies on the functionality implemented in the ports, i.e. within Action $Set_1$ and Action $Set_2$ of Fig. 2. Thus, CFM can be applied for SDN VLANs too. Furthermore, CFM can be used for the virtual overlay networks as well, e.g. between VMs in a DC. The SDN Controller can instantiate and set up the operation (e.g. time period for monitoring) in the ports that need to be involved. The proper CFM actions out of Action $Set_1$ and Action $Set_2$ can be then automatically performed on the OAM packets. That is, hybrid networks make the full blown, proven and already used Ethernet OAM available for SDN too, thus providing the OAM tools at each Layer 2 virtual network overlay.

Protection switching state machines based on CFM are also specified by [1] for point-to-point VLANs, hence fast protection switching is available for SDN too. Therefore, the SDN control can instantiate protection switching as well if needed, e.g. for a carrier-grade service. In addition, a hybrid network can leverage further features specified by the standard. For example the features specified by the IEEE 802.1 Data Center Bridging (DCB) working group are essential in Cloud deployments relying on Ethernet, e.g. support for VM migration [6].

Direct collaboration between SDN and distributed control is required to realize the full potential of the architecture. First, in addition to manipulating the forwarding behavior, SDN has to be able to set up and control the functionality that has distributed components, e.g. OAM in order to use it for SDN traffic. Second, the SDN Controller should be aware of the topology, the service assignments and the load in some form to exercise effective control. This can be achieved by retrieving the necessary information from the distributed control. Then, upon request for establishment of a service, the SDN Controller is able to select the control to be used based on the requirements of the service and/or on the actual state of the network. For instance, the default shortest path is satisfactory for certain services, while other services may require full path control, and may also require OAM and protection switching. For the latter type of services, the SDN Controller programs the forwarding path, sets up and initiates the operation of OAM and protection switching based on the exact service requirements. That is, the level of interaction between the SDN Controller and distributed control depends on the service requirements; therefore, the SDN touch points for service establishment may vary as well.

Overall, taking advantage of the hybrid approach allows meeting key network requirements, e.g. OAM. SDN control can implement auto-provisioning based on the auto-discovery of topology and services provided by SPB. In addition, QoS can be enhanced by the proper assignment of services to the appropriate control, i.e. to SDN or SPB. Further work is going on in the form of P802.1Qca, which aims to better exploit the potential in hybrid networking based on IS-IS.

The most important aspect is that by relying on existing packet and tunneling formats, the hybrid approach enables re-using existing chips and avoids the need for complete replacement of the data plane, i.e. it is at most software upgrade to the existing devices.

## VI. A NETWORK EXAMPLE

After exploring the networking principles in the previous sections, let us investigate the operation. A hybrid network example comprising nodes supporting both SDN and SPB is discussed in the following. The example PBB network is shown in Fig. 7. The I-SIDs provide overlay Virtual Networks (VN) to S-VIDs in the example. Two new virtual networks: $VN_1$ and $VN_2$ are just being created in the example by the SDN Controller.

Let us assume that the multipoint-to-multipoint $VN_1$ has no special requirements, thus the SDN Controller decides to use shortest paths for $VN_1$. Therefore, the SDN Controller only touches the end points, i.e. it programs the proper associations in the Edge Bridges (EB) supporting $VN_1$. Thus, $S-VID_{11}$ is associated with I-SID$_1$, which is then associated with B-VID$_1$ in EB$_1$, EB$_2$ and EB$_3$. B-VID$_1$ is allocated to the SPBM MSTI, therefore, it is a non-SDN VLAN controlled by SPB. The rest of actions for the establishment of VN$_1$ are then performed by the distributed link state SPB; the SDN Controller has no further task. SPB provides the service discovery, thus the Core Bridges (CB) become aware of that EB$_1$, EB$_2$ and EB$_3$ are member of the virtual network identified by I-SID$_1$. Therefore, SPB populates the forwarding tables in the CBs to establish the multipoint-to-multipoint B- VID$_1$ transport tunnel for I-SID$_1$, which then provides the connectivity service to S-VLAN$_{11}$. If a new end point is required for a service due to e.g. a station (VM) movement, then after setting the proper associations at the required EB, SPB automatically establishes the connectivity, thus supporting station (VM) migration.

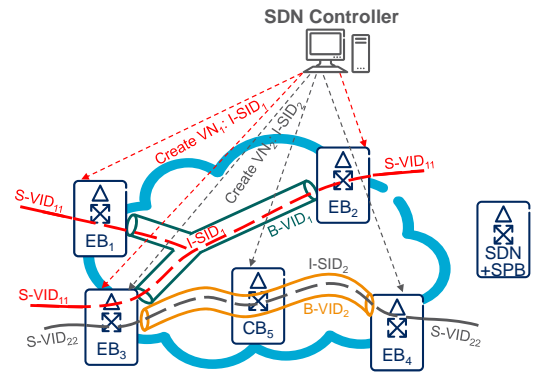

Fig. 7. A PBB network example

Let us assume that based on its requirements, VN$_2$ needs full path control, which may deviate from the shortest path. Therefore, the SDN Controller has to program the forwarding at all bridges along the path in addition to performing the proper associations at EB$_3$ and EB$_4$. Thus, S-VID$_{22}$ is associated with I-SID$_2$, which is then mapped to B-VID$_2$ in the EBs. B-VID$_2$ is an SDN VLAN because it is allocated to the Ext-MSTI. Therefore, the distributed control does not touch this service.

An interface is required between SPB and the SDN Controller to allow the SDN Controller to retrieve the link state database of SPB, e.g., from one of the bridges. Thus, the SDN Controller can rely on SPB to discover the physical topology; furthermore, the service discovery provided by SPB can be also used by the SDN Controller, at least for verification.

VII. SUMMARY

This paper has shown that the basic design principles of Ethernet bridging are in-line with SDN and today's network requirements. We set forward three key principles for such SDN architectures:

*1)* The use of an existing, data plane model (Ethernet). This includes features that require complex processing in the data plane, such as OAM or protection switching. Such features are difficult to implement solely using centralized SDN.

*2)* Co-existence with and reliance on distributed control plane for useful features, such as topology discovery and path setup, where applicable. Such co-existence can happen by the two control planes controlling different layers or side-by-side controlling different parts of the traffic (e.g., separated by the VLAN space).

*3)* The coordinated communication between the distributed and SDN control planes. This can be useful, for example, for the SDN control plane to learn the topology already discovered by the distributed control plane. Such communication could also enable the SDN control plane to react to topology changes and to adjust the parameters of the distributed control plane, when needed.

The resulting SDN architecture has natural limitations, especially due to the first point above. We argue, however, that for early SDN systems the benefits of readily available features outweigh the limitations.

Furthermore, Ethernet provides a good basis for future extensions and for the evolution of SDN, e.g. along the lines of the third point above. Future work may involve the research on the interface between the control planes.